\documentclass[aps,prx,twocolumn,letterpaper,tighten,float,nofootinbib]{revtex4}
\usepackage{graphicx}
\usepackage{subfigure}
\usepackage{latexsym}
\usepackage{amsmath,amssymb}
\usepackage{bm} 
\usepackage{color}
\usepackage{stackrel}
\usepackage{epsfig}

\definecolor{myblue}{rgb}{.93, .93, 1}

\newcommand{\bsub}{\begin{subequations}}
\newcommand{\esub}{\end{subequations}}

\begin{document}

\title{Mott glass from localization and confinement}
\author{Yang-Zhi~Chou}\email{YangZhi.Chou@colorado.edu} \author{Rahul~M.~Nandkishore} \author{Leo~Radzihovsky}

\affiliation{Department of Physics and Center for Theory of Quantum
	Matter, University of Colorado Boulder, Boulder, Colorado 80309,
	USA} \date{\today}

\date{\today}

\begin{abstract}
  We study a system of fermions in one spatial dimension with linearly
  confining interactions and short-range disorder.  We focus on the
  zero temperature properties of this system, which we characterize
  using bosonization and the Gaussian variational method. We compute
  the static compressibility and ac conductivity, and thereby
  demonstrate that the system is incompressible, but exhibits gapless
  optical conductivity. This corresponds to a ``Mott glass'' state,
  distinct from an Anderson and a fully gapped Mott insulator, arising
  due to the interplay of disorder and charge confinement. We argue
  that this Mott glass phenomenology should persist to non-zero
  temperatures.
 \end{abstract}

\maketitle

\section{Introduction}

Rich phenomena arising from the interplay of disorder and interactions in
quantum many-body systems have been at the center of considerable
excitement over the past decade, particularly with the advent of
many-body localization (MBL) \cite{Anderson1958,
  Basko06,Gornyi2005,Nandkishore2014}, a phenomenon whereby disordered
interacting systems can exhibit ergodicity breaking and fail to
equilibrate even at infinite times. While most theory in this field
has been formulated for systems with short-range interactions, and
long-range interactions are typically expected to suppress
localization \cite{Levitov1990,Burin2006,Yao2014}, a recent work
\cite{Nandkishore2017} introduced a model of one-dimensional (1D)
linearly interacting fermions -- the celebrated Schwinger model
\cite{Schwinger1962,Coleman1976,Wolf1985,Fischler1979} -- with the
additional ingredient of quenched disorder, argued to exhibit many-body 
localization despite its long-range interactions.

Here we study the ground state and low-energy properties of the
disordered Schwinger model. While 1D non-interacting
fermions in a random potentials localize, forming a gapless
compressible Anderson insulator, a clean 1D system with
Coulomb interaction exhibits charge confinement, with a fully gapped
ground state. What are the ground state and low-energy properties when
both of these ingredients are present, namely that of disordered
linearly interacting fermions in one dimension?

Using bosonization and Gaussian variational method (GVM)
\cite{Mezard1991,Giamarchi1996}, here we explore the zero-temperature
properties of this model and demonstrate that it realizes a distinct
phase of matter, a ``Mott glass'' \cite{Orignac1999,Giamarchi2001},
that is characterized by a hard gap in compressibility, but not in
optical conductivity, i.e., it exhibits a vanishing compressibility
and a finite ac conductivity down to zero frequency.  The hard gap in
compressibility arises due to confinement of charged excitations,
while the absence of a hard gap in optical conductivity is due to the
existence of random localized dipole excitations down to zero
energy. Unlike previous explorations of such phenomena
\cite{Orignac1999,Giamarchi2001,Weichman2008,Vojta2016}, Mott glass in
the disordered Schwinger model is driven by the interplay of disorder
and confinement from long-range interactions, and does not require a
commensurate periodic potential \cite{Orignac1999}. The model evades
the compelling arguments against the Mott glass phase advanced in
Ref.~\onlinecite{Nattermann2007} by way of its long-range
interactions (a loophole that was anticipated in
Ref.~\onlinecite{Nattermann2007}).
 
While linearly confining interactions do not naturally arise in the
solid state, the Schwinger model has been proposed
\cite{Rico2014,Kuhn2014,Notarnicola2015,Yang2016} and realized
\cite{Martinez2016} in synthetic quantum systems.  Furthermore, it has
been extensively explored via numerical simulations
\cite{Byrnes2002,Rico2014,Buyens2014,Banuls2015,Saito2015,Banuls2016,Buyens2016,Brenes2017,Kasper2017}. The
ideas advanced herein therefore admit near term tests both in numerics
and in experiments with synthetic quantum matter.

The article is organized as follows.  In Sec.~II, we introduce the
disordered Schwinger model and formulate its bosonized form.  We then
briefly review the GVM analysis in Sec.~III. Readers familiar with
these details may skip directly to Sec.~IV, where the optical
conductivity and static compressibility are calculated using the GVM,
and Mott glass physics is demonstrated.  We discuss the implications
of our results and conclude in Sec.~V.

\section{Model}\label{Sec:Model}

\begin{figure}[b]
	\centering
	\includegraphics[width=0.35\textwidth]{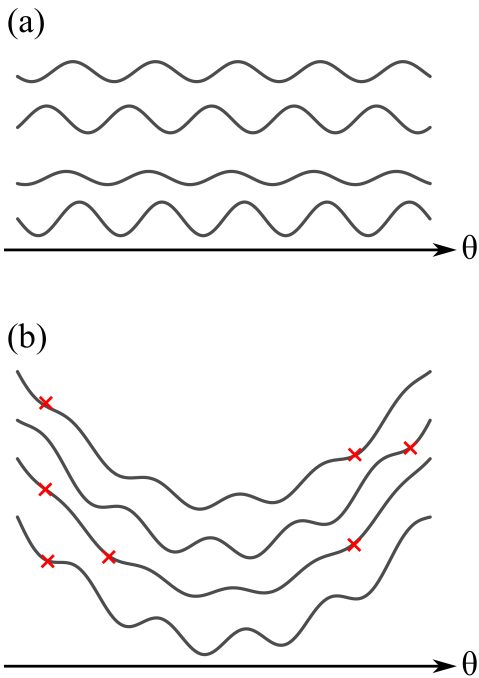}
        \caption {An ensemble of random boson potential in the
          disordered boson problems, with curves (offset for clarity)
          corresponding to different spatial points.  (a) The
          Giamarchi-Schulz model exhibits a spatially random
          potential, that is periodic in the phonon displacement
          $\theta$, and is therefore characterized by an infinite
          number of minima. (b) In contrast, because of the fermionic
          long-range interactions a weakly disordered Schwinger model
          generically exhibits a unique minimum selected by the
          overall quadratic phonon potential. (Occasionally, the phonon potential 
          develops doubly degenerate minima.) The red crosses indicate
          the local minima removed by the long-range fermionic
          interaction and the associated bosonic quadratic (in
          $\theta$) potential. The resulting ground state properties
          are qualitatively distinct from those of the conventional
          Bose glass, as we discuss in the main text.}
	\label{Fig:theta_x}
\end{figure}

We study a disordered Schwinger model \cite{Schwinger1962,Coleman1976}
that describes electrons interacting via a dynamical gauge field in
one spatial and one temporal dimension (1+1D).  The gauge field
induces a linearly-confining electron potential, corresponding to the
Fourier transform of $1/k^2$, screened by a positive uniform ionic
background (jellium model) and with the uniform translational zero
mode $k=0$ suppressed by the boundary conditions.  Integrating out the
dynamical gauge field in the disorder-free Schwinger model leads to a
strongly-interacting Hamiltonian $H_0+H_{\text{int}}$, where
\begin{align}
H_0=&v\int_x\left[R^{\dagger}\left(-i\partial_x R\right)-L^{\dagger}\left(-i\partial_x L\right)\right],\\
\label{Eq:H_int}H_{\text{int}}=&-e^2\int_{x,x'} \,\rho_0(x)\left|x-x'\right|\rho_{0}(x'),
\end{align}
$\rho_0=R^{\dagger}R+L^{\dagger}L$ is the long-wavelength component of
the density at average incommensurate filling, written in terms of the
right ($R$) and left ($L$) moving chiral fermion, and above we defined
$\int_x\equiv\int dx$.
Utilizing standard bosonization \cite{Giamarchi_Book,Shankar_Book}, the
disorder-free Schwinger model is equivalently formulated in terms of
an imaginary time path-integral with Euclidean action for a
phonon-like field $\theta$ given by
\begin{align}
\mathcal{S}_0=&\int_{\tau,x}\left[
\frac{1}{2\pi vK}
\left((\partial_{\tau}\theta)^2+v^2\left(\partial_x\theta\right)^2
\right)+ \frac{2e^2}{\pi^2}\theta^2\right],
\label{Eq:S_0}
\end{align}
with the charge gap due to confinement characterized by a plasma
frequency $\omega_p=2e\sqrt{vK/\pi}$ and a Luttinger parameter $K$,
which accounts for short-range interactions.

The key additional ingredient that we include \cite{Nandkishore2017} is
the random impurities, modeled by a short-range correlated disorder
potential. The latter can be expressed in terms of the long-wavelength
forward-scattering ($H_{\text{imp},f}$) and the short-scale
backscattering ($H_{\text{imp},b}$) parts \cite{Giamarchi1988}, given
by
\begin{align}
H_{\text{imp},f}=&\int_x\,\eta(x)\rho_0(x),\\
H_{\text{imp},b}=&\int_x\,\left[\xi(x) L^{\dagger}R+\xi^*(x)R^{\dagger}L\right].
\end{align}
The corresponding random potentials, $\eta(x)$ (forward-scattering
scalar potential) and $\xi(x)$ (back-scattering potential that is a
random mass in the electronic representation
\cite{Kogut1975,Byrnes2002}) are respectively real and complex
zero-mean Gaussian random fields, characterized by
\begin{align}
\overline{\eta(x)\,\eta(x')}=&\Delta_f\delta(x-x'),\,\,\,\overline{\xi(x)\,\xi^*(x')}=\Delta\delta(x-x'),\\
\overline{\eta(x)\,\xi(x')}=&\overline{\xi(x)\,\xi(x')}=0,
\end{align}
where $\overline{\mathcal{O}}$ denotes disorder average of
$\mathcal{O}$.

The forward-scattering component of disorder bosonizes to
$\int_{\tau,x} \,\eta(x)\,\frac{1}{\pi}\partial_x\theta$ in Euclidean action, and (since
it is linear in $\theta$) can be safely eliminated from the action by
a shift in $\theta$ linear in $x$ (see discussions in
Appendix.~\ref{App:scalar}), and for our purposes can thus be safely
neglected as in the conventional localization problem
\cite{Giamarchi1988}.
In contrast, the back-scattering disorder plays a crucial role and
leads to localization. Integrating over disorder using a replica
``trick'' \cite{Edwards1975} (in equilibrium equivalent to the Keldysh
path integral), generates a short-range disorder component of the
replicated action,
\begin{align}\label{Eq:S_dis_b}
\mathcal{S}_{\text{dis}}=-\tilde{\Delta}
\sum_{a,b}\int_{x,\tau,\tau'}
\cos\left[2\theta_a(\tau,x)-2\theta_b(\tau',x)\right],
\end{align}
where $\tilde{\Delta}=\Delta/(2\pi^2\alpha^2)$, $a$ and $b$ are
replica indices, and $\alpha$ is the microscopic ultra-violet length
scale, set for example by the disorder correlation length or the
underlying lattice constant.

Because long-range Coulomb interaction strongly suppresses charge
fluctuations (see Fig.~\ref{Fig:theta_x}), gapping out $\theta$, in contrast to randomly pinned
acoustic systems \cite{Fisher1989} and short-range interacting
disordered electrons (as studied by e.g., Giamarchi-Schulz
\cite{Giamarchi1988}) the back-scattering disorder in the Schwinger model
is strongly relevant for all values of parameters,
\begin{align}\label{Eq:RG_dis_b}
\frac{d\tilde{\Delta}}{d l}=3\tilde{\Delta}.
\end{align}

Our interest is in the ground state and low-energy excitations of the
disordered Schwinger model, where fermions are confined by the linear
potential, and bosonic excitations are localized by
disorder \cite{Nandkishore2017}.  The strong relevance of
disorder thus requires a nonperturbative treatment in $\Delta$, to
which we turn next.

\section{Gaussian Variational Method}
We now analyze the low-energy properties of the model defined by
$\mathcal{S} = \mathcal{S}_0 + \mathcal{S}_{\text{dis}}$
[Eqs.~(\ref{Eq:S_0}) and (\ref{Eq:S_dis_b})], utilizing a
nonperturbative but generally uncontrolled GVM
\cite{Mezard1991,Giamarchi1996} . The basic idea is to
approximate the Schwinger nonlinear action by the ``best'' harmonic
action, with optimized variational parameters determined by the
minimum of the variational ground state energy.  The resulting
quadratic variational action then allows a computation of physical
observables, with our focus here on the optical conductivity and
static compressibility, that characterize the disordered Schwinger
ground state and its low-energy excitations.

The GVM is known to capture the basic low-temperature properties of
Bose glass in the Giamarchi-Schulz model \cite{Giamarchi1996}.
Because the Schwinger model is gapped by long-range interactions,
cutting off long-scales (that are otherwise challenging to handle) by
the inverse of the plasma frequency, $1/\sqrt{\omega_p}$, we in fact
expect the GVM to be both qualitatively and quantitatively accurate
for the problem at hand, at least for weak disorder.

To this end, we consider a general imaginary time action $\mathcal{S}$
which can be separated into two parts,
$\mathcal{S}=\mathcal{S}_{\text{var}}+\delta \mathcal{S}$, where
$\mathcal{S}_{\text{var}}$ is a variational quadratic action.  $\delta
\mathcal{S}=\mathcal{S}-\mathcal{S}_{\text{var}}$ is a perturbation
around $\mathcal{S}_{\text{var}}$.  The partition function is formally
expressed as
\begin{align}
Z=\int \mathcal{D}\theta\,e^{-\mathcal{S}_{\text{var}}[\theta]-\delta\mathcal{S}[\theta]}=Z_0\left\langle e^{-\delta\mathcal{S}[\theta]}\right\rangle_{\text{var}},
\end{align}
where $\langle\mathcal{O}\rangle_{\text{var}}$ denotes the expectation
value of $\mathcal{O}$ with respect to $\mathcal{S}_{\text{var}}$.
Accordingly, the free energy is given by
\begin{align}
F=&-T\ln Z_{\text{var}} -T\ln\left[\left\langle e^{-\delta\mathcal{S}}\right\rangle_{\text{var}}\right]\\
\le &-T\ln Z_{\text{var}}-T\left\langle\delta\mathcal{S}\right\rangle_{\text{var}}\equiv F_{\text{var}},
\end{align}
where $F_{\text{var}}$ is a variational free energy functional to
leading order in $\delta\mathcal{S}$, that by convexity of the
exponential function is a strict upper bound for the actual free
energy $F$ \cite{Feynman,Chaikin_Lubensky}.  Although this can be extended to an
improved variational free energy upper bound as a cumulant expansion
in $\delta\mathcal{S}$, here we limit our analysis to above lowest
order, as it is sufficient for our purposes here.  The optimal
$\mathcal{S}_{\text{var}}$ upper bound is set by minimizing the
variational free energy $F_{\text{var}}$.

For the disordered Schwinger model, we consider the replicated
disorder-averaged action,
$\mathcal{S}=\mathcal{S}_{\text{var}}+\mathcal{S}_{\text{dis}}$. The
inter- and intra- replica correlation functions need to be treated as
independent functions.  Even though only the intra-replica response
function is directly related to physical observables, correlation
functions are determined by an inverse of the replica matrix kernel
and thus depend on all of its components. The general variational
quadratic action is given by:
\begin{align}
\nonumber\mathcal{S}_{\text{var}}=&\frac{1}{2\beta L}\sum_a\sum_{\omega_n,k}G^{-1}_{aa}(\omega_n,k)\theta_a(-\omega_n,-k)\theta_a(\omega_n,k)\\
&+\frac{1}{2\beta L}\sum_{a,b,(a\neq b)}\sum_{\omega_n,k}G^{-1}_{ab}(\omega_n,k)\theta_a(-\omega_n,-k)\theta_b(\omega_n,k),
\end{align}
where $a$, $b$ denote replica indices, $\beta$ is the inverse
temperature, and $L$ is the system size.  The intra-replica $(G_{aa})$
and inter-replica $(G_{a\neq b})$ Green functions are independent
variational parameters.

With this set up, the variational free energy
$F_{var}=F_0+F_1+F_{\text{dis}}$ is then formally given by,
\begin{align}
F_0=&-\frac{1}{2\beta}\sum_a\sum_{\omega_n,k}\ln\left[G(\omega_n,k)\right]_{aa},\\
F_1=&
\frac{1}{2v\pi K\beta}\sum_a\sum_{\omega_n,k}\left(\omega_n^2+v^2k^2+\frac{4e^2vK}{\pi}\right)G_{aa}(\omega_n,k),\\
\label{Eq:F_dis}F_{\text{dis}}=&
-\tilde{\Delta} \int_{\tau,x}
\left[
\sum_{a,b}
e^{
	-2\left\langle 
	\left(\theta_a(\tau,x)-\theta_b(0,x)\right)^2
	\right\rangle_{\text{var}}}\right].
\end{align}
where $F_0$ is the free energy corresponding to the harmonic
variational action $\mathcal{S}_{\text{var}}$, $F_1$ and
$F_{\text{dis}}$ come from $-T\langle
\left(\mathcal{S}_0-\mathcal{S}_{\text{var}}\right)\rangle_{\text{var}}$
and $-T\langle \mathcal{S}_{\text{dis}}\rangle_{\text{var}}$
contributions, respectively, and we have dropped an unimportant
additive constant. The bosonic correlator in Eq.~(\ref{Eq:F_dis}) is
straightforwardly computed to be given by
\begin{align}
\nonumber&\left\langle 
\left(\theta_a(\tau,x)-\theta_b(0,x)\right)^2
\right\rangle_{\text{var}}\\
=&\frac{1}{\beta L}\sum_{\omega_n,k}\left[G_{aa}(\omega_n,k)+G_{bb}(\omega_n,k)-2\cos\left(\omega_n\tau\right)G_{ab}(\omega_n,k)\right].
\end{align}

We then carry out a functional derivative of $F_{var}$ with respect to
the variational parameters, $G_{ab}(\omega_n,k)$. The saddle point
equation is thereby given by
\begin{align}
\nonumber G^{-1}_{ab}=&\left[\frac{1}{v\pi K}
\left(\omega_n^2+v^2k^2\right)+\frac{4e^2}{\pi^2}\right]\delta_{ab}\\
\nonumber&+4\tilde{\Delta}\sum_{cd}\int_{\tau} \bigg\{\!
\left[\delta_{ab}\delta_{ac}+\delta_{ab}\delta_{ad}-2\delta_{ac}\delta_{bd}\cos\left(\omega_n\tau\right)\right]\\
&\label{Eq:SPEqn_G}
\hspace{0.4cm}\times
e^{-\frac{2}{\beta L}
\sum\limits_{\nu_n,k}\left[G_{cc}(\nu_n,q)+G_{dd}(\nu_n,q)-2\cos\left(\nu_n\tau\right)G_{cd}(\nu_n,q)\right]}\bigg\}.
\end{align}
and determines the optimum value of the variational parameters.  The
explicit derivation is standard but lengthy, with details found in the
literature \cite{Giamarchi1996,Orignac1999,Giamarchi2001} and specific
to the disordered Schwinger model in Appendix.~\ref{App:SPE}.

Before turning to the computation of the Green function and the
associated physical predictions, we highlight key technical components
of the analysis. In the replica formalism, the inter-replica
correlations are time-independent \cite{Giamarchi1996}.  Consequently,
$G^{-1}_{a\neq b}(\omega_n,k)$ is only non-zero at $\omega_n=0$.
Another important detail is that the optimum variational solution is
given by a replica symmetry broken structure, with $G^{-1}_{a b}$ for
$a\neq b$ given by a hierarchical
structure \cite{Mezard1991,Giamarchi1996}.  For 1D disordered fermions,
it has been demonstrated that a one-step replica symmetry broken
solution\footnote{This means that there are only two distinct
  components of the inter-replica Green functions.} is sufficient, and
the marginal stability condition is adopted \cite{Giamarchi1996}.

For the disordered Schwinger model, the variational ansatz is the same
as that previously used in Refs. \onlinecite{Orignac1999} and \onlinecite{Giamarchi2001}
for their putative, substrate-driven Mott glass \cite{Nattermann2007},
except that here the boson mass is (not a variational parameter but
is) physical, determined by the plasma frequency associated by the
confining Coulomb interaction.  We separate the intra-replica Green
functions into the finite-frequency ($\omega_n\neq 0$) and
zero-frequency ($\omega_n= 0$) components. The former is given by a
diagonal matrix in the replica space. The latter is derived by
inverting a hierarchical matrix in the replica space
\cite{Mezard1991}. The Green function is therefore non-analytic at
zero frequency. For finite frequencies, the intra-replica correlation
function is given by
\begin{align}
\label{Eq:G_loc_fw}G_{aa}(\omega_n\neq 0,k)
=&\frac{1}{\frac{1}{\pi v K}\left(\omega_n^2+v^2k^2\right)+\frac{4e^2}{\pi^2}+\Sigma+I(\omega_n)},
\end{align}
where $\Sigma+I(\omega_n)$ is the bosonic self energy, with $\Sigma$
its zero-frequency component, and $I(\omega_n)$ determines the
frequency dependence of the self energy, and is crucial for response
functions like optical conductivity. $I(\omega_n)$ has the following
important asymptotic behaviors,
\begin{align}\label{Eq:I_omega_n_asym}
I(\omega_n)\sim\begin{cases}
\frac{2}{\sqrt{3}}\sqrt{\frac{1}{v\pi K}\left(\Sigma+\frac{4e^2}{\pi^2}\right)}|\omega_n|,& \text{for }\omega_n\rightarrow 0,\\[2mm]
2\left(\Sigma+\frac{4e^2}{\pi^2}\right)\left[1-\frac{\sqrt{v\pi K\left(\Sigma+\frac{4e^2}{\pi^2}\right)}}{|\omega_n|}\right],& \text{for }\omega_n \rightarrow \infty.
\end{cases}
\end{align}
We obtain the full frequency dependence of $I(\omega_n)$ by solving
Eq.~(\ref{Eq:SC_I_MS}) derived from GVM \cite{Giamarchi1996}.

For $\omega_n=0$, the intra-replica Green function takes a different
form from Eq.~(\ref{Eq:G_loc_fw}) and is given by
\begin{align}
\label{Eq:G_loc_zw}G_{aa}(\omega_n=0,k)=\frac{1/u_c}{\frac{v}{\pi K}k^2+\frac{4e^2}{\pi^2}}
+\frac{1-1/u_c}
{\frac{v}{\pi K}k^2+\Sigma+\frac{4e^2}{\pi^2}},
\end{align}
where $u_c\in [0,1]$ [determined by Eq.~(\ref{Eq:Sigma_uc})] is a
parameter of the one-step replica symmetry broken ansatz.

We thus find that the variational solution of the disordered Schwinger
model displays a form (finite boson mass) similar to that of the
putative Mott glass state, proposed in
Refs.~\onlinecite{Orignac1999} and \onlinecite{Giamarchi2001}. However, here the
ever-present charge confinement ensures that the boson mass is always
nonzero, and thus the disordered Schwinger model does not exhibit a
transition to an Anderson insulator (vanishing mass) or a fully gapped
insulator ($\Sigma=0$) in the thermodynamic limit. As we will see in
the next section, for all ranges of parameters at zero temperature it
displays the phenomenology of and therefore realizes the Mott glass
phase proposed in Ref.~\onlinecite{Orignac1999}, but without requiring
a commensurate lattice potential, and evading arguments in
Ref.~\onlinecite{Nattermann2007}.

\section{Response Functions}

The universal behavior of physical observables can be used to define
and distinguish qualitatively distinct phases. For example, combining
the results of optical conductivity and static compressibility, one
can distinguish metals, gapped insulators, and Anderson insulators.
Here we study the optical conductivity and the compressibility of the
disordered Schwinger model, and demonstrate below that indeed they
exhibit qualitative behavior consistent with the Mott glass phase
envisioned in Refs.~\onlinecite{Orignac1999} and \onlinecite{Giamarchi2001}.  We will
then also discuss the implications for the low-temperature states
based on the Mott glass phenomenology.

\begin{figure}[b]
\centering
\includegraphics[width=0.4\textwidth]{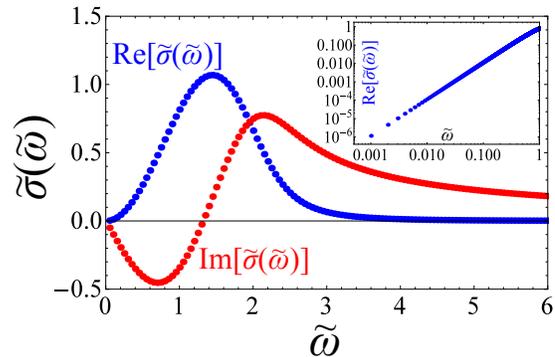}
\caption{Dimensionless optical conductivity in the ground state of the
  disordered Schwinger model given by Eq.~(\ref{Eq:optical_cond}). The
  dimensionless quantities are defined by
  $\tilde{\sigma}(\tilde\omega)=\pi^2\sigma(\omega)/\sqrt{\frac{v\pi K}{\Sigma'}}$ and
  $\tilde\omega=\omega \sqrt{\frac{1}{v\pi
      K\Sigma'}}$. $\Sigma'=\Sigma+\frac{4e^2}{\pi^2}$. The blue (red)
  dots denote the real (imaginary) part of the $\tilde\sigma(\tilde\omega)$.  The
  real part of the optical conductivity does not contain a hard gap
  but displays a power law $\text{Re}\left[\tilde{\sigma}(\tilde\omega)\right]\propto \tilde\omega^2$
  behavior in the low frequency limit, as illustrated in the
  inset. The optical conductivity is consistent with that of a
  localized insulator. Inset: The low frequency real part of the
  conductivity, displays $\omega^2$ behavior.}
\label{Fig:optical_cond}
\end{figure}

\subsection{Optical Conductivity}

To compute linear optical conductivity, we use Kubo formula and
analytic continuation to real frequency of the intra-replica
correlation functions. The optical conductivity is thus given by
\begin{align}\label{Eq:ac_cond}
\sigma(\omega)=\frac{1}{\pi^2}\frac{i}{\omega+i\delta}\left[\omega_n^2\left\langle\theta\theta\right\rangle(\omega_n,k=0)\right]_{i\omega_n\rightarrow \omega+i\delta},
\end{align}
with $\delta\rightarrow 0^+$. The associated $\theta$ phonon
correlator above is straightforwardly computed using the optimized
quadratic variational action based on GVM. Using this correlator and
Eqs.~(\ref{Eq:G_loc_fw}) and (\ref{Eq:ac_cond}), the optical
conductivity is then given by
\begin{align}\label{Eq:optical_cond}
\sigma(\omega)=&\frac{1}{\pi^2}\frac{i}{\omega+i\delta}\left[\frac{-\omega^2}{\frac{-1}{v\pi K}\omega^2+\frac{4e^2}{\pi^2}+\Sigma+I(-i\omega)}\right]\\
\label{Eq:optical_cond2}=&\frac{-i\omega}{\pi^2}
\frac{\frac{-\omega^2}{v\pi K}+\Sigma'+\text{Re}[I(-i\omega)]-i\text{Im}[I(-i\omega)]}{\left(\frac{-\omega^2}{v\pi K}+\Sigma'+\text{Re}[I(-i\omega)]\right)^2+\left(\text{Im}[I(-i\omega)]\right)^2},
\end{align}
where $\Sigma'=\Sigma+\frac{4e^2}{\pi^2}$. The finite frequency
dependence of $\sigma(\omega)$ is determined by $I(-i\omega)$. The
full frequency profile of the optical conductivity is plotted in
Fig.~\ref{Fig:optical_cond}.  We focus on low and high frequency
asymptotic dependences characterizing the Mott glass phase, that can
be worked out via Eqs.~(\ref{Eq:I_omega_n_asym}) and
(\ref{Eq:optical_cond2}).

In the low frequency limit, the real and imaginary part of
conductivity give
\begin{align}
\text{Re}[\sigma(\omega)]\approx
&\frac{1}{\pi^2}\frac{\sqrt{\frac{1}{v\pi K}}}{\left(\Sigma+\frac{4e^2}{\pi^2}\right)^{3/2}}\frac{2}{\sqrt{3}}\omega^2,\\
\text{Im}[\sigma(\omega)]\approx&\frac{-1}{\pi^2}\frac{\omega}{\Sigma+\frac{4e^2}{\pi^2}}.
\end{align}
The low temperature conductivity behavior is consistent with the
previous GVM analysis for Bose glass models in one dimension
\cite{Giamarchi1996,Orignac1999,Giamarchi2001} and coincides with the
states deep in many-body localized phase \cite{Gopalakrishnan2015}
(for a different reason, presumably). It is consistent with the
gapless compressible Anderson insulator, with the optical conductivity
characterized by a low-frequency power law down to zero
frequency \cite{Mott1968}. In one dimension, it is given by $\omega^2$ upto
logarithmic corrections \cite{Mott1968,Berezinskii1974}.  Indeed the
GVM gives $\omega^2$ behavior, but cannot resolve logarithmic
correction in the low frequency limit. We emphasize that the low
frequency $\omega^2$ dependence is determined by $I(-i\omega)$ in
Eq.~(\ref{Eq:optical_cond2}). In contrast, a fully gapped Mott or band
insulator below a gap is characterized by $I(-i\omega)=0$, displaying
a hard gap in the optical conductivity.

In the high frequency limit (still in a localized regime), the real
and imaginary parts of conductivity give
\begin{align}
\text{Re}[\sigma(\omega)]\approx&
\frac{2\sqrt{\pi}(vK)^{5/2}}{\omega^4}\left(\Sigma+\frac{4e^2}{\pi^2}\right)^{3/2},\\
\text{Im}[\sigma(\omega)]\approx&\frac{vK}{\pi}\frac{1}{\omega}.
\end{align}
In the Giamarchi-Schulz model, this tail in the real part of
conductivity reproduces the perturbative result\footnote{For
  sufficiently high frequencies, electrons oscillate within a length
  scale smaller than the localization length, justifying a
  perturbative approach.}, $\text{Re}[\sigma(\omega)]\sim
\omega^{-4+2K}$ \cite{Giamarchi_Book}, by considering finite
temperature and weak disorder in GVM \cite{Giamarchi1996}.  In the
disordered Schwinger model, the strongly relevant renormalization
group flow of the backscattering disorder [given by
Eq.~(\ref{Eq:RG_dis_b})] is independent of the Luttinger parameter,
thereby effectively corresponding to $K=0$. We therefore predict the
$\omega^{-4}$ high frequency tail in the non-zero (low) temperature
limit.

\subsection{Compressibility}
A complementary characterization of the state is via the static
compressibility, that is nonzero for a compressible Anderson
insulator, and zero for an incompressible gapped Mott insulator.

Within GVM, the compressibility of the disordered Schwinger model is
straightforwardly computed from the static density-density correlation
function,
\begin{align}\label{Eq:D_k}
D(k)=&\frac{k^2}{\pi^2}\left\langle\theta(\omega=0,-k)\,\theta(\omega=0,k)\right\rangle\\
=&k^2\left[\frac{1/u_c}{\frac{v}{\pi K}k^2+\frac{4e^2}{\pi^2}}
+\frac{1-1/u_c}
{\frac{v}{\pi K}k^2+\Sigma+\frac{4e^2}{\pi^2}}\right],
\end{align}
where we have used the zero-frequency Green function given by
Eq.~(\ref{Eq:G_loc_zw}) from GVM. As advertised in the Introduction,
we thus find that the static compressibility of the disordered
Schwinger model, $\chi_s\propto\lim\limits_{k\rightarrow 0}D(k)=0$,
vanishes, displaying a charge gap associated with confinement.
Namely, $\chi_s$ would remain be finite in the absence of $4e^2/\pi^2$
in Eq.~(\ref{Eq:D_k}). This result appears to be in conflict with our
results for optical conductivity (which contains no hard gap). In the
next subsection we will discuss the differences between the two
response functions, arguing that this is a consistent characterization
of the Mott glass phase, intermediate between and sharply distinct
from a fully compressible Anderson and a fully incompressible Mott
insulator.

\subsection{Mott Glass Phenomenology}

The ground state of the disordered Schwinger model simultaneously
exhibits localization and confinement.  Above we have demonstrated
above, its low-frequency optical conductivity displays $\omega^2$
behavior, a hallmark of a localized insulator in one dimension
\cite{Mott1968,Berezinskii1974}.  On the other hand, we found that its
static compressibility vanishes, characteristic of a fully gapped
incompressible insulator.  This unconventional ground state is thus a
Mott glass \cite{Orignac1999,Giamarchi2001,Weichman2008},
characterized by gapped single particle and gapless particle-hole
excitations. In the presence of disorder, the particle-hole
excitations appear at arbitrary low energies but remain localized in
space. \cite{Giamarchi2001,Nandkishore2017}.  These localized
excitons dominate the optical conductivity at low frequencies, while
the static compressibility vanishes due to the absence of low energy
charged (single particle) excitations.

This intermediate Mott glass phenomenology naturally characterizes the
ground state of disordered fermions with confinement of charges driven
by long-range interaction. We conjecture that Mott glass may also
extend to non-zero temperature states (putatively MBL
\cite{Nandkishore2017}) as we now discuss.

Firstly, confinement is a property of the entire spectrum, not just
the ground state.  We therefore expect that the static density-density
correlation function $D(k\rightarrow 0)$ [given by Eq.~(\ref{Eq:D_k})]
is vanishingly small even at the non-zero energy-density many-body
states. Although it is difficult to infer much about the finite
energy density MBL states based on the GVM results, it is generally
believed that the optical conductivity in MBL gives $\omega^{\alpha}$
where $\alpha$ is a continuously varying exponent between 1 and 2
\cite{Gopalakrishnan2015}. For states deep inside the MBL phase,
$\omega^2$ behavior is expected. Given the arguments of Ref.~\onlinecite{Nandkishore2017}, 
indicating that the disordered Schwinger
model is many-body localized at non-zero energy density (at least
insofar as states can be many-body localized in the continuum
\cite{Aleinercontinuum, 2dcontinuum, gornyicontinuum}), we conjecture
that Mott glass phenomenology of a disordered Schwinger model also
extends to non-zero temperature.

Finally, it is interesting to consider the effect of a uniform
background electric field, $E_0$, that can be included by adding $\sim
\int_{x,\tau} e E_0 x \rho_0(x)$ to the action. Bosonizing $\rho_0(x)
= \frac{1}{\pi} \partial_x \theta$ and integrating by parts, it is
clear that an electric field appears as $-\int_{x,\tau} e E_0 \theta$,
and can be shifted away (along with the forward scattering), leaving
the disorder-averaged action $\mathcal{S}_{0} +
\mathcal{S}_{\text{dis}}$ unchanged. 
Therefore, in contrast to the conventional Schwinger model studied by
Coleman \cite{Coleman1976}, consistent with
Imry-Wortis \cite{ImryWortis} arguments, in a disordered Schwinger
model we do not expect a uniform electric field to induce a phase
transition.

\section{Discussion and Conclusion}

We have studied a disordered Schwinger model, describing
one-dimensional, long-range interacting relativistic fermions in the
presence of a random potential.  We find that the model exhibits a
localized state, despite its long-range interactions.  We study its
properties and compute its optical conductivity and static
compressibility within the Gaussian replica variation analysis.  We
find that the system shows an incompressible localized glass state,
with low-frequency conductivity scaling with $\omega^2$. We thus show
that such a system indeed displays properties akin to a putative Mott glass 
state, previously proposed in a different context in the
literature\cite{Orignac1999,Giamarchi2001}. 

By way of long-range confining interactions the present system
sidesteps forceful arguments against the existence of Mott glass,
advanced in Ref.~\onlinecite{Nattermann2007}.  Our results indicate
that the Mott glass phenomenology (gapped single-particle and gapless
localized particle-hole excitations) is a natural consequence of the
simultaneous presence of localization and confinement. Whether this
phenomenology persists in more complicated disordered confined systems
(perhaps in higher dimensions) is an intriguing question for future
work. 


We furthermore conjecture that Mott glass remains stable at non-zero
temperatures, up to small corrections associated with fragility of MBL
in the continuum \cite{Aleinercontinuum, 2dcontinuum,
  gornyicontinuum}.

While the linearly confining potential does not naturally arise in
conventional solid state materials, the Schwinger model may be
realized and studied in the synthetic quantum many-body systems
\cite{Rico2014,Kuhn2014,Notarnicola2015,Yang2016,Martinez2016,Kasper2017}.
Numerical simulations
\cite{Byrnes2002,Rico2014,Buyens2014,Banuls2015,Saito2015,Banuls2016,Buyens2016,Brenes2017}
also provide a route to explore this interesting model and to test our
predictions for its phenomenology.

\section*{Acknowledgements}

We thank Venkitesh Ayyar, Tom DeGrand, Masanori Hanada, Jed Pixley,
Michael Pretko, and Hong-Yi Xie for useful discussions.  This work is
supported in part by a Simons Investigator award from the Simons
Foundation, and NSF grant no. DMR-1001240 (Y.-Z.C. and L.R.), and in
part by the Army Research Office under Grant Number W911NF-17-1-0482
(Y.-Z.C. and R.N.). The views and conclusions contained in this
document are those of the authors and should not be interpreted as
representing the official policies, either expressed or implied, of
the Army Research Office or the U.S. Government. The U.S. Government
is authorized to reproduce and distribute reprints for Government
purposes notwithstanding any copyright notation herein. R.N. also
acknowledges the support of the Alfred P. Sloan foundation through a
Sloan Research Fellowship.

\appendix

\section{Spatially-dependent Scalar Potential}\label{App:scalar}

In this Appendix, we demonstrate how forward scattering potential in
$\mathcal{S}_0+\mathcal{S}_{\eta}$ can be eliminated by a simple
time-independent shift of the phonon field
$\theta(x,\tau)\rightarrow\theta(x,\tau)+\zeta(x)$. Under this
transformation the action becomes $\mathcal{S}'$
\begin{align}
\nonumber\mathcal{S}'=&\mathcal{S}_0
+\int d\tau dx\frac{2v^2}{2 v\pi K}\left(\partial_x\theta\right)\left(\partial_x\zeta\right)\\
&+\int d\tau dx\frac{2M^2}{2\pi}\theta\zeta+\int d\tau dx \,\eta(x)\frac{1}{\pi}\partial_x\theta,
\end{align}
where we have dropped unimportant, $\theta$-independent constant. We
then require a vanishing of the terms linear in $\theta$, ensured by
by a choice of the shift $\zeta(x)$ field satisfying a saddle point equation
\begin{align}\label{Eq:Saddel_point_Eq}
\partial_x^2\zeta-\frac{KM^2}{v}\zeta=-\frac{K}{v}\partial_x\eta(x).
\end{align}
Under the physical condition of $\zeta(\pm \infty)=0$, the solution is
then simply given by
\begin{align}
\label{Eq:zeta_x}\zeta(x)
=&\int dx' \frac{e^{-M\sqrt{K/v}|x-x'|}}{2M\sqrt{K/v}}\left[-\frac{K}{v}\partial_{x'}\eta(x')\right].
\end{align}

\section{Variational Free Energy and Saddle Point Equations}\label{App:SPE}

In this appendix, for completeness we sketch the solution of
Eq.~(\ref{Eq:SPEqn_G}), following the analysis in
Ref.~\onlinecite{Giamarchi1996}.

To begin, we define the self-energy $\chi_{ab}$ via
\begin{align}
G^{-1}_{ab}(\omega_n,k)=&\left[\frac{\omega_n^2+v^2k^2}{v\pi K}
+\frac{M^2}{\pi}\right]\delta_{ab}-\chi_{ab}(\omega_n,k),
\end{align}
where $M^2=4e^2/\pi$.  It is also useful to define a quantity
$G_C^{-1}=\sum_bG^{-1}_{ab}$, which satisfies the saddle point equation,
\begin{align}
\nonumber G_C^{-1}(\omega_n,k)=&\frac{1}{v\pi K}
\left(\omega_n^2+v^2k^2\right)+\frac{M^2}{\pi}\\
\nonumber&+8\tilde{\Delta}\!\int d\tau\Bigg\{\!\left[1-\cos\left(\omega_n\tau\right)\right]
\\
\label{Eq:SPE_GVM_1}&\hspace{1.5cm}\times\left[
e^{-2B_{aa}(\tau)}
+\!\sum\limits_{b,(b\neq a)}\!e^{-2B_{ab}(\tau)}
\right]\!\Bigg\},\\
\label{Eq:SPE_GVM_2}\chi_{a\neq b}(\omega_n,k)=&8\tilde\Delta\int d\tau\cos\left(\omega_n\tau\right)
e^{-2B_{ab}(\tau)},
\end{align}
where 
\begin{align}
B_{aa}(\tau)=&\frac{2}{\beta L}\sum_{\omega_n,k}\left[G_{aa}(\omega_n,k)-\cos(\omega_n\tau)G_{aa}(\omega_n,k)\right],\\
B_{a\neq b}(\tau)=&\frac{2}{\beta L}\sum_{\omega_n,k}\left[G_{aa}(\omega_n,k)-\cos(\omega_n\tau)G_{a b}(\omega_n,k)\right]\\
=&\frac{2}{\beta L}\sum_{\omega_n,k}\left[G_{aa}(\omega_n,k)-\delta_{\omega_n,0}G_{a b}(0,k)\right].
\end{align}
Above we have used the symmetry $G_{ab}=G_{ba}$ and constancy of the
diagonal elements.  We also note that, $B_{a\neq b}$ is independent of
$\tau$ \cite{Giamarchi1996}.


In order to compute the saddle point equations in the required
zero-replica limit, we adopt Parisi's parametrization
\cite{Mezard1991} as follows:
\begin{align}
A_{aa}\rightarrow \tilde{A},\,\,\, A_{a\neq b}\rightarrow A(u),
\end{align}
where $\tilde{A}$ is the intra-replica element and $u\in[0,1]$ is a
continuous parameter that encodes the inter-replica structure in
$A(u)$.  Specifically, we consider one-step replica symmetry broken
ansatz which corresponds to $A(u<u_c)=A_0$ and $A(u\ge
u_c)=A(u_c)$. $u_c$, describing the break point of $u$ is also a
variational parameter in GVM. We also adopt the algebraic rules of the
hierarchical matrices in Ref.~\onlinecite{Mezard1991}.

The saddle point equations [Eqs.~(\ref{Eq:SPE_GVM_1}) and
(\ref{Eq:SPE_GVM_2})] in the zero-replica limit become to
\begin{align}
\nonumber G_C^{-1}(\omega_n,k)=&
\frac{1}{v\pi K}
\left(\omega_n^2+v^2k^2\right)+\frac{M^2}{\pi}\\
\nonumber &+8\tilde\Delta\int d\tau\left[1-\cos\left(\omega_n\tau\right)\right]\\
\label{Eq:SPE_GVM_ZR_1}&\hspace{1cm}\times\left[
e^{-2 \tilde{B}(\tau)}
-\int_{0}^{1}du \,e^{-2 B(u)}
\right],\\
\nonumber\chi(\omega_n;u)
=&8\tilde\Delta\int d\tau\cos\left(\omega_n\tau\right)
e^{-2B(u)}\\
\label{Eq:SPE_GVM_ZR_2}=&8\tilde\Delta\beta \delta_{\omega_n,0}
e^{-2B(u)}\equiv \delta_{\omega_n,0}\chi(u)
\end{align}
where
\begin{align}
\tilde{B}(\tau)=&\frac{2}{\beta L}
\sum_{\nu_n,q}\left[\tilde{G}(\nu_n,q)-\cos\left(\nu_n\tau\right)\tilde{G}(\nu_n,q)\right],\\
B(u)=&\frac{2}{\beta L}
\sum_{\nu_n,q}\left[\tilde{G}(\nu_n,q)-\delta_{\nu_n,0}G(\nu_n,q;u)\right].
\end{align}
We note that the summation over inter-replica elements turns into an
integration over $u$ with an overall minus sign due to zero-replica
limit.  The inter-replica correlations vanish for non-zero Matsubara
frequencies \cite{Giamarchi1996}. Therefore, we can simply invert the
Green function $G^{-1}_C(\omega_n\neq 0,k)=[G_C(\omega_n\neq
0,k)]^{-1}$.  We express $\tilde{B}(\tau)$ and $B(u)$ as follows:
\begin{align}
\label{Eq:B_t}\tilde{B}(\tau)=&\frac{2}{\beta L}
\sum_{\nu_n,q}\left[1-\cos\left(\nu_n\tau\right)\right]G_C(\nu_n,q),\\
\nonumber B(u)=&\frac{2}{\beta L}
\sum_{\nu_n\neq 0,q}G_C(\nu_n,q)\\
\label{Eq:B}&+\frac{2}{\beta L}\sum_{q}\left[\tilde{G}(\nu_n=0,q)-G(\nu_n=0,q;u)\right].
\end{align}

For one-step replica symmetry broken ansatz, we consider \cite{Giamarchi1996,Orignac1999,Giamarchi2001}
\begin{align}
\chi(u)=\begin{cases}
\chi(u_c), & \text{ for }u_c\ge u,\\
0, & \text{ for }u_c< u.
\end{cases}
\end{align}
Correspondingly,
\begin{align}
B(u)=\begin{cases}
B, & \text{ for }u_c\ge u,\\
\infty, & \text{ for }u_c< u.
\end{cases}
\end{align}

The Green functions [$G_C(\omega_n,k)$, $\tilde{G}(\omega_n,k)$, and
$G(\omega_n,k;u)$ ] depend on $\chi(u)$. We introduce self-energy
parameters $\Sigma$ and $I(\omega_n)$ to encode the interacting Green
functions,
\begin{align}
\nonumber G_C^{-1}(\omega_n,k)=&\frac{1}{\pi v K}\left(\omega_n^2+v^2k^2\right)+\frac{M^2}{\pi}\\
\label{Eq:G_C}&+\left(1-\delta_{\omega_n,0}\right)\Sigma+I(\omega_n),
\end{align}
where
\begin{align}
\label{Eq:I_omega}I(\omega_n)=&8\tilde\Delta\int d\tau\left[1-\cos(\omega_n\tau)\right]\left[e^{-2\tilde{B}(\tau)}-e^{-2B}\right],\\
\Sigma=&=u_c\chi(u_c)=8\tilde\Delta u_c\beta e^{-2\alpha^2B}.
\end{align}
The structure of $G_C^{-1}$ encodes translational invariance after
disorder average, with $I(\omega_n)$ vanishing as $\omega_n$ goes to
zero. In addition to $G_C^{-1}$, we also need to examine the zero
frequency Green functions. In particular,
\begin{align}
\label{Eq:G-Gu}
\tilde{G}(0,k)-G(0,k;u)=\begin{cases}
\tilde{G}(0,k), &\text{ for }u<u_c,\\[1mm]
\frac{1}{\frac{v}{\pi K}k^2+\frac{M^2}{\pi}+\Sigma},&\text{ for }u\ge u_c,
\end{cases}
\end{align}
where $G(0,k)$ is given by Eq.~(\ref{Eq:G_loc_zw}) and we have used the inversion formula of hierarchical matrices \cite{Mezard1991}.

With Eqs.~(\ref{Eq:B_t}), (\ref{Eq:B}), and (\ref{Eq:G-Gu}), we obtain
\begin{align}
B-\tilde{B}(\tau)
=&\frac{2}{\beta L}\sum_{\nu_n,q}\cos(\nu_n\tau)G_C(\nu,q)
\end{align}
and
$\lim\limits_{\tau\rightarrow\infty}\lim\limits_{\beta\rightarrow\infty}\tilde{B}(\tau)=B$. 

For determining the explicit frequency dependence of $I(\omega_n)$, we
expand Eq.~(\ref{Eq:I_omega}) to leading order of $B-\tilde{B}(\tau)$,
treating it as a small parameter for a sufficiently large $\tau$. The
self-consistent equation for $I(\omega_n)$ then reduces to
\begin{align}\label{Eq:SC_I_1}
I(\omega_n)
=&\frac{2\Sigma}{ u_c  \beta}\sqrt{\frac{\pi K}{v}}
\left[\frac{1}{\sqrt{\Sigma'}}-\frac{1}{\sqrt{\left[\Sigma'+I(\omega_n)+\frac{\omega_n^2}{\pi v K}\right]}}\right],
\end{align}
where $\Sigma'=\Sigma+\frac{M^2}{\pi}$. To close the equations, one
needs to obtain the expression of $\Sigma$ and $u_c$ as
well. Following Giamarchi and Le Doussal, we use marginal stability
\cite{Giamarchi1996}, that determines the solution for 1D interacting
fermion problems. We first assume that $I(\omega_n)\approx
c_1|\omega_n|$ for small frequencies. The existence of a solution in
Eq.~(\ref{Eq:I_omega}) can be expressed as
\begin{align}
&c_1|\omega_n|=\frac{4\Sigma}{u_c}\frac{1}{\beta}\int_{-\infty}^{\infty}\frac{dq}{2\pi}\frac{c_1|\omega_n|}{\left[\frac{1}{\pi v K}\left(v^2q^2\right)+\Sigma'\right]^2}\;,\\
\label{Eq:Sigma_uc}\rightarrow\;\;& \frac{\Sigma}{u_c}\frac{1}{\beta}\sqrt{\frac{\pi K}{v}}=\Sigma'^{3/2}\;.
\end{align}
With this, Eqs.~(\ref{Eq:Sigma_uc}) and (\ref{Eq:SC_I_1}) lead to a
simple self-consistent equation as follows,
\begin{align}\label{Eq:SC_I_MS}
I(\omega_n)=2\Sigma'^{3/2}\left[\frac{1}{\sqrt{\Sigma'}}-\frac{1}{\sqrt{\left[\Sigma'+I(\omega_n)+\frac{\omega_n^2}{\pi v K}\right]}}\right]\;.
\end{align}

We define $\tilde{I}=I/\Sigma'$, $z=\sqrt{(1/v\pi
  K)(\omega_n^2/\Sigma')}$, and $\Sigma'=\Sigma+\frac{M^2}{\pi}$.
Equation (\ref{Eq:SC_I_MS}) is simplified by
\begin{align}
&\tilde{I}(z)=2\left(1-\frac{1}{\sqrt{1+\tilde{I}(z)+z^2}}\right),
\end{align}
that gives $\tilde{I}(z)$, which determines the optical conductivity
discussed in the main text, with limiting forms given by
\begin{align}
\tilde{I}(z)\sim\begin{cases}
\frac{2}{\sqrt{3}}z,& \text{ for }z\sim 0,\\
2-\frac{2}{z},& \text{ for }z \rightarrow \infty.
\end{cases}
\end{align}




\end{document}